\documentclass[final]{article}
\usepackage{idris2}

\usepackage{relsize}

\usepackage{tikz}
\usetikzlibrary{shapes.geometric}

\usepackage{catchfilebetweentags}
\input{robust-catch}

\usepackage{acronym}
\usepackage{xspace}
\usepackage[todonotes={obeyFinal},commandnameprefix=always]{changes}
\usepackage{jfdm-plt}
\usepackage{velo}
\usepackage{environ}
\setuptodonotes{inline}

\usepackage{authblk}

\pdfoutput=1

\title{Type Theory as a Language Workbench} 

\author[1]{Jan {de Muijnck-Hughes}}
\author[2]{Guillaume {Allais}}
\author[2]{Edwin {Brady}}

\affil[1]{University of Glasgow}
\affil[2]{University of St Andrews}

\newcommand{\syntaxtypes}{
\[\begin{array}{lcl}
  \ty{t}{\Type}
  & \Coloneqq
  & \TyNat \\
  & \fpAlt
  & \TyBool \\
  & \fpAlt
  & \typeFuncIntro{}
\end{array}\]}

\newcommand{\syntaxcontexts}{
\[\begin{array}{lcl}
  \ty{\Gamma}{\Context}
  & \Coloneqq
  & \epsilon \\
  & \fpAlt
  & \Gamma,\, \ty{x}{t}
\end{array}\]}

\newcommand{\varRule}{
  $\Gamma \ni \ty{x}{a}$
}

\newcommand{\varZero}{
  \[
  \infer{ }{\Gamma \,, \ty{x}{a} \ni \ty{x}{a}}
  \]
}

\newcommand{\varSuc}{
  \[
  \infer{\Gamma \ni \ty{x}{a}}{\Gamma \,, \ty{y}{b} \ni \ty{x}{a}}
  \]
}

\newcommand{\inferenceRule}{
  $\Gamma \vdash \ty{t}{a}$
}

\newcommand{\inferenceZero}{
  \[
  \infer{ }{\Gamma \vdash \ty{\exprZero}{\TyNat}}
  \]
}

\newcommand{\inferenceVar}{
  \[
  \infer{\Gamma \ni \ty{x}{a}}{\Gamma \vdash \ty{x}{a}}
  \]
}

\newcommand{\inferenceInc}{
  \[
  \infer{\Gamma \vdash \ty{n}{\TyNat}
    }{\Gamma \vdash \ty{\exprInc{n}}{\TyNat}}
  \]
}

\newcommand{\inferenceApp}{
  \[
  \infer{\Gamma \vdash \ty{f}{\TyFunc{a}{b}}
        \\ \Gamma \vdash \ty{t}{a}
    }{\Gamma \vdash \ty{\exprApp{f}{t}}{b}}
  \]
}

\newcommand{\inferenceFunc}{
  \[
  \infer{\Gamma,\, \ty{x}{a} \vdash \ty{t}{b}
    }{\Gamma \vdash \ty{\exprLam{t}}{\TyFunc{a}{b}}}
  \]
}

\newcommand{\holeexamplegraph}{
\begin{tikzpicture}

\draw (0,0)   circle (.4cm) node[align=center] {$\lambda a.$};
\draw (2.5,0) circle (.4cm) node[align=center] {\$};

\draw (5,.75) circle (.4cm) node[align=center] {$\lambda x.$};
\draw[dashed] (7.5,.75) circle (.4cm) node[align=center] {$?h$};

\draw (5,-.75) circle (.4cm) node[align=center] {$\lambda y.$};
\draw[dashed] (7.5,-.75) circle (.4cm) node[align=center] {$?h$};

\draw [->] (.4,0)     to [out=0,in=180]  (2.1,0);
\draw [->] (2.8,.25)  to [out=45,in=180] (4.6,.75);
\draw [->] (2.8,-.25) to [out=-45,in=180] (4.6,-.75);

\draw [->] (5.4,.75)  to [out=0,in=180] (7.1,.75);
\draw [->] (5.4,-.75) to [out=0,in=180] (7.1,-.75);

\end{tikzpicture}}

\newcommand{\cseexamplegraph}{
\begin{tikzpicture}
\draw (1.6,0)    circle (.4cm) node[align=center] {\$};

\draw (3.75,.75) circle (.4cm) node[align=center] {$\lambda x.$};
\draw (6.90,.75) node[draw,dashed,regular polygon, regular polygon sides=3, shape border rotate=90]
      {\phantom{b}$t$\phantom{g}};

\draw (3.75,-.75) circle (.4cm) node[align=center] {$\lambda a.$};
\draw (5.90,-.75) circle (.4cm) node[align=center] {$\lambda b.$};
\draw (9,-.75) node[draw,dashed,regular polygon, regular polygon sides=3, shape border rotate=90]
      {\phantom{b}$t$\phantom{g}};

\draw [->] (1.9,.27)  to [out=45,in=180] (3.35,.75);
\draw [->] (1.9,-.27) to [out=-45,in=180] (3.35,-.75);

\draw [->] (4.15,-.75) to [out=0,in=180]  (5.5,-.75);
\draw [->] (4.15,.75)  to [out=0,in=180]  (5.9,.75);

\draw [->] (6.3,-.75) to [out=0,in=180]  (8,-.75);

\end{tikzpicture}}

\newcommand{\codebruijnexamplegraph}{
\begin{tikzpicture}
\draw[draw=blue   ,thick] (0,0) circle (.3cm) node[align=center] {$\lambda{}.$};
\draw[draw=magenta,thick] (2,0) circle (.3cm) node[align=center] {$\lambda{}.$};
\draw[draw=orange ,thick] (4,0) circle (.3cm) node[align=center] {$\lambda{}.$};

\draw (6,0) circle (.3cm) node[align=center] {\$};

\draw (8,1)    circle (.3cm) node[align=center] {\$};
\draw[fill=blue]   (10,1.5) +(-.3cm,-.3cm) rectangle +(.3cm,.3cm);
\draw[fill=orange] (10,.5)  +(-.3cm,-.3cm) rectangle +(.3cm,.3cm);

\draw (8,-1)    circle (.3cm) node[align=center] {\$};
\draw[fill=magenta] (10,-.5)  +(-.3cm,-.3cm) rectangle +(.3cm,.3cm);
\draw[fill=orange]  (10,-1.5) +(-.3cm,-.3cm) rectangle +(.3cm,.3cm);

\draw [->] (0.3,0)   to [out=0,in=180] node[above]{$\color{blue}{\bullet}$} (1.7,0);
\draw [->] (2.3,0)   to [out=0,in=180] node[above]{$\color{blue}{\bullet}\color{magenta}{\bullet}$} (3.7,0);
\draw [->] (4.3,0)   to [out=0,in=180] node[above]{$\color{blue}{\bullet}\color{magenta}{\bullet}\color{orange}{\bullet}$} (5.7,0);

\draw [->] (6.2,-.22)  to [out=-45,in=180] node[below left]{$\color{blue}{\circ}\color{magenta}{\bullet}\color{orange}{\bullet}$} (7.7,-1);
\draw [->] (6.2,.22)   to [out=45,in=180] node[above left]{$\color{blue}{\bullet}\color{magenta}{\circ}\color{orange}{\bullet}$} (7.7,1);
\draw [->] (8.2,1.22)  to [out=45,in=180] node[above]{$\color{blue}{\bullet}\color{orange}{\circ}$} (9.7,1.5);
\draw [->] (8.2,.78)   to [out=-45,in=180] node[below]{$\color{blue}{\circ}\color{orange}{\bullet}$} (9.7,.5);
\draw [->] (8.2,-.78)  to [out=45,in=180] node[above]{$\color{magenta}{\bullet}\color{orange}{\circ}$} (9.7,-.5);
\draw [->] (8.2,-1.22) to [out=-45,in=180] node[below]{$\color{magenta}{\circ}\color{orange}{\bullet}$} (9.7,-1.5);
\end{tikzpicture}
}

\acrodef{dsl}[DSL]{Domain Specific Language}
\acrodef{edsl}[EDSL]{Embedded Domain Specific Language}
\acrodef{stlc}[STLC]{Simply-Typed Lambda Calculus}
\acrodef{cse}[CSE]{Common Sub-Expression Elimination}
\acrodef{ir}[IR]{Intermediate Representation}
\acrodef{repl}[REPL]{Read Eval Print Loop}
\acrodef{lsp}[LSP]{Language Server Protocol}

\makeatletter
\newcommand*{\org@overidelabel}{}
\let\org@overridelabel\@verridelabel
\@ifpackagelater{acronym}{2015/03/21}{
  \renewcommand*{\@verridelabel}[1]{%
    \@bsphack
    \protected@write\@auxout{}{\string\AC@undonewlabel{#1@cref}}%
    \org@overridelabel{#1}%
    \@esphack
  }%
}{
  \renewcommand*{\@verridelabel}[1]{%
    \@bsphack
    \protected@write\@auxout{}{\string\undonewlabel{#1@cref}}%
    \org@overridelabel{#1}%
    \@esphack
  }%
}
\makeatother

\definechangesauthor[color=orange,name={Jan}]{jfdm}
\definechangesauthor[color=green, name={Guillaume}]{gallais}
\definechangesauthor[color=red, name={Edwin}]{edwin}

\newcommand{\Velo}{V{\'e}lo\xspace}
\newcommand{\Idris}{Idris~2}
\newcommand{\DeBruijn}{De~Bruijn}

\fvset
{ xleftmargin=0.5\parindent
, commandchars=\\\{\}
}

\DefineVerbatimEnvironment%
  {VerbatimInline}
  {Verbatim}
  {xleftmargin=0.5\parindent
  , frame=leftline
  }
\DefineVerbatimEnvironment%
  {VerbatimInlineNumbered}
  {Verbatim}
  {xleftmargin=\parindent
    ,fontsize=\smaller
    ,numbers=left,numbersep=2pt
    ,frame=lines
  }

\NewEnviron{centertight}[1][0.5em]
{\vspace{#1}
 {\centering\BODY}
 \vspace{#1}
}

\usepackage{hyperref}
\usepackage{cleveref}

\begin{document}

\maketitle


\begin{abstract}
  Language Workbenches offer language designers an expressive environment in which to create their \acp*{dsl}.
  Similarly, research into mechanised meta-theory has shown how dependently typed languages provide expressive environments to formalise and study \acsp*{dsl} and their meta-theoretical properties.
  But can we claim that dependently typed languages qualify as language workbenches?
  We argue yes!

  We have developed an exemplar \acs*{dsl} called \Velo{} that showcases not only dependently typed techniques to realise and manipulate \acp*{ir}, but that dependently typed languages make fine language workbenches.
  \Velo{} is a simple verified language with well-typed holes and comes with a complete compiler pipeline: parser, elaborator, \acs*{repl}, evaluator, and compiler passes.
  Specifically, we describe our design choices for well-typed \acs*{ir} design that includes support for well-typed holes, how \acf*{cse} is achieved in a well-typed setting, and how the mechanised type-soundness proof for \Velo{} is the source of the evaluator.
\end{abstract}


%
%
%
%

\todo{Run spellcheck}
\todo{Use Katla to typeset Idris 2 code}

\section{Introduction}
\label{sec:introduction}

\emph{Language Workbenches}, such as
Spoofax~\cite{DBLP:journals/software/WachsmuthKV14},
offer language designers an expressive environment in which to design,
implement, and deploy their \Acp{dsl}~\cite{hudak1996building}.
Principally speaking a language workbench~\cite{DBLP:conf/sle/ErdwegSVBBCGHKLKMPPSSSVVVWW13}
is a tool that supports:
description of a language's \emph{notation}---how we present a language's concrete syntax to users;
implementation of a language's \emph{semantics}---how we realise the language's behaviour;
and user interaction through an \emph{editor}.
Outside of these core criteria, various language workbenches
support language validation, testing, and composition.

Concurrently, the mechanised meta-theory research
programme~\cite{DBLP:conf/tphol/AydemirBFFPSVWWZ05,DBLP:journals/jfp/AbelAHPMSS19}
has seen a wealth of tools and techniques being developed
by the programming languages theory community.
In particular, dependently typed languages such as
\Idris{}~\cite{DBLP:conf/ecoop/Brady21},
Agda~\cite{DBLP:conf/afp/Norell08},
and Coq~\cite{the_coq_development_team_2022_5846982}
have been widely used to formalise \acp{dsl}, and study their
meta-theoretical properties.
Dependent types allow types to depend on values --- that is,
types are first class --- and provide an expressive environment
in which to reason about, and write, our programs.
Efforts using dependently typed languages range from
studying specific core calculi~\cite{10.1145/3093333.3009866,DBLP:conf/cpp/RouvoetPKV20,DBLP:conf/mpc/ChapmanKNW19}
to building generic reasoning frameworks~\cite{DBLP:conf/cpp/StarkSK19,DBLP:journals/jfp/AllaisACMM21}.
These mechanised software verification projects, however, typically stop short
of building the frontend that would let users run these verified
language implementations.
If our verified language implementations type check, we might as well ship them too!
By becoming its own implementation language, \Idris{} has successfully
demonstrated that this is not an inescapable fate~\cite{DBLP:conf/ecoop/Brady21}.
But can we now claim that dependently typed languages qualify as
language workbenches?

\Velo{}\footnote{A reproducible artifact, and the source code, has been provided as supplementary material.} is a minimal functional language that we have realised in \Idris{}
to showcase dependently typed techniques to implement and manipulate \acp{ir}.
This paper introduces \Velo{} but, most of all, seeks to show that
dependently typed languages make fine language workbenches.
We address both the core criteria and some optional extensions
highlighted by the language workbench challenge~\cite{DBLP:conf/sle/ErdwegSVBBCGHKLKMPPSSSVVVWW13} for what constitutes a language workbench.
Although not all of the optional criteria
are met by dependently typed languages, we are convinced that
with some additional engineering (taking advantage of existing work,
for example Quickchick~\cite{DBLP:journals/pacmpl/LampropoulosPP18})
more optional criteria can be satisfied.

Another key tenet in language workbenches, such as Spoofax,
is the \emph{ease} with which languages can be created.
To that same degree, we have developed a series
of reusable modules
that captures
functionality common to many languages,
thereby reducing the \emph{boilerplate} required when creating \acp{edsl} in \Idris{}.

Although we have made an effort to make dependently typed
programming accessible in our presentation, more introductory
material is available for the interested reader~\cite{plfa22.08,brady17:_type_driven_devel_idris}.

\todo{Add link to DOI-backed artifact}


\section{Introducing Velo}
\label{sec:velo}

The design behind \Velo{} is purposefully unsurprising:
it is the \ac{stlc} extended with let-bindings,
booleans and their conjunction,
and natural numbers and their addition.
To promote the idea of interactive editing \Velo{} also supports well-typed holes.
Below we show an example \Velo{} program, which contains a multiply used hole, and an extract from the \acs*{repl}
session that lists the current set of holes.

\begin{center}
  \begin{minipage}[t]{0.55\linewidth}
\begin{Verbatim}
   let b = false
in let double
         = (fun x : nat => (add x x))
in let x = (double ?hole)
in         (double ?hole)
\end{Verbatim}
\end{minipage}
\hfill
  \begin{minipage}[t]{0.35\linewidth}
    \begin{Verbatim}
Velo> :holes
b : Bool
double : Nat -> Nat
----------
?hole : Nat
\end{Verbatim}
\end{minipage}

\end{center}

The featherweight language design of \Velo{} helps us
showcase better how we can use dependently typed languages
as language workbenches~\cite{DBLP:journals/toplas/IgarashiPW01}.
Regardless of language complexity, \Velo{} is nonetheless a
complete language with a standard compiler pipeline, and \acs*{repl}.
A \ac{dsl} captures the language's concrete syntax, and a parser turns \ac{dsl} instances into raw unchecked terms.
Bidirectional type checking keeps type annotations to a minimum in the concrete syntax, and helps to better elaborate raw un-typed terms into a set of well-typed \acp{ir}:
\IdrisType{Holey} to support well-scoped typed holes;
and
\IdrisType{Terms} the core representation that captures our language's abstract syntax.
We present interesting aspects of our \ac{ir} design in \Cref{sec:design}.
Further, elaboration performs standard desugarings that e.g. turns let-bindings into function application thus reducing the size of our core.
From the core representation we also provide well-scoped \ac{cse} using co-\DeBruijn{} indexing (\Cref{sec:compiler-pass}), and we provide a verified evaluator to reduce terms to values (\Cref{sec:semantics}).

\todo{Show an example high-level trace of the output?}


\section{Language Design}
\label{sec:design}

We begin our discussion by detailing the key design rationale on
realising the static semantics of \Velo{} within \Idris{}.
We have opted to give \Velo{} an external concrete syntax (a \ac{dsl})
in which users can write their programs.
With dependently typed languages we can also capture
the abstract syntax and its static semantics as an intrinsically
scoped and typed \ac{edsl}
directly within the host language~\cite{Augustsson1999edt}.
That is to say that the data structure is designed in such a way that
we can only represent well scoped and well typed terms and, correspondingly,
that our scope- and type- checking passes are guaranteed to have rejected
invalid user inputs.
To keep the exposition concise, we focus on a core subset of the
language. The interested reader can find the whole definition
in the accompanying material\todo{Add DOI}.

\textbf{Types} are usually introduced using their context free grammar.
We present it here on the left-hand side, it gives users the choice between
two base types (\TyNat, and \TyBool) and a type former for function types
(\TyFunc{\cdot}{\cdot}).
On the right hand side, we give their internal representation as an inductive
type in \Idris{}.%
\footnote{
Throughout this article, the \Idris{} code snippets are
automatically rendered using a semantic highlighter.
Keywords are typeset in \IdrisKeyword{bold},
types in \IdrisType{blue},
data constructors in \IdrisData{red},
function definitions in \IdrisFunction{green},
bound variables in \IdrisBound{purple},
and comments in \IdrisComment{grey}.
}

\begin{centertight}
\begin{minipage}{0.45\textwidth}
\syntaxtypes
\end{minipage}\hfill
\begin{minipage}{0.45\textwidth}
\ExecuteMetaData[Code/MiniVelo.idr.tex]{TyDef}
\end{minipage}
\end{centertight}

\textbf{Contexts} can be similarly given by a context free grammar:
a context is either empty ($\epsilon$), or an existing context ($\Gamma$)
extended on the right with a new type assignment (\ty{x}{t}) using a comma.
In \Idris{}, we will adopt a nameless representation and so we represent
these contexts by using a \IdrisData{SnocList} of types
(i.e.\ lists that grow on the right).
Note that the \Idris{} compiler automatically supports sugar for lists and
snoc lists: \IdrisData{[1,2,3]} represents a list counting up from
\IdrisData{1} to \IdrisData{3} while \IdrisData{[<1,2,3]} is its snoc list
pendant counting down.
In particular \IdrisData{[<]} denotes the empty snoc list also known as \IdrisData{Lin}.

\begin{centertight}
\begin{minipage}{0.35\textwidth}
\syntaxcontexts
\end{minipage}\hfill
\begin{minipage}{0.55\textwidth}
\ExecuteMetaData[Code/MiniVelo.idr.tex]{SnocListDef}
\end{minipage}
\end{centertight}

\todo{play with spaces, reread descriptions}

\textbf{Typing Judgements} are given by relations, and encoded in
\Idris{} using inductive families, a generalisation of inductive
types~\cite{DBLP:journals/fac/Dybjer94}.
Each rule will become a constructor for the family, and so every
proof \inferenceRule{} will correspond to a term $t$ of type
(\texttt{Term} $\Gamma$ $a$).
On the left hand side we present two judgements: context membership
and a typing judgement, and on the right we have the corresponding
inductive family declarations.

\begin{centertight}
\begin{minipage}{0.10\textwidth}
\varRule
\inferenceRule
\end{minipage}\hfill
\begin{minipage}{0.80\textwidth}
\ExecuteMetaData[Code/MiniVelo.idr.tex]{ElemTermDecl}
\end{minipage}
\end{centertight}

We leave the definition of \IdrisType{Elem} to the next section,
focusing instead on \IdrisType{Term}.
The most basic of typing rules are axioms, they have no premise
and are mapped to constructors with no argument.
We use \Idris{} comments (\IdrisComment{\KatlaDash{}\KatlaDash})
to format our constructor's type in such
a way that they resemble the corresponding inference rule.
Here we show the rule stating that $0$ is a natural number and
its translation as the \IdrisData{Zero} constructor.

\begin{centertight}
\begin{minipage}{0.45\textwidth}
\inferenceZero
\end{minipage}\hfill
\begin{minipage}{0.45\textwidth}
\ExecuteMetaData[Code/MiniVelo.idr.tex]{inferenceZero}
\end{minipage}
\end{centertight}

Then come typing rules with a single premise which is not a subderivation
of the relation itself.
They are mapped to constructors with a single argument.
Here we show the typing rule for variables: given a proof that we have a
variable of type $a$ somewhere in the context, we can build a term of type
$a$ in said context.

\begin{centertight}
\begin{minipage}{0.45\textwidth}
\inferenceVar
\end{minipage}\hfill
\begin{minipage}{0.45\textwidth}
\ExecuteMetaData[Code/MiniVelo.idr.tex]{inferenceVar}
\end{minipage}
\end{centertight}

Next, we have typing rules with a single premise which is a subderivation.
They are mapped to constructors with a single argument of the inductive family
representing the subderivation.
Here we show the typing rule for successor: provided that we are given
a natural number in a given context, its successor is also a natural
number in the same context.

\begin{centertight}
\begin{minipage}{0.45\textwidth}
\inferenceInc
\end{minipage}\hfill
\begin{minipage}{0.45\textwidth}
\ExecuteMetaData[Code/MiniVelo.idr.tex]{inferenceInc}
\end{minipage}
\end{centertight}

Similarly, rules with two premises are translated to constructors
with two arguments, one for each subderivation.
Here we present the typing rule for application nodes: provided that
the function has a function type, and the argument has a type matching
the function's domain, the application has a type corresponding to the
function's codomain.
Note that the context $\Gamma$ is the same across the whole rule and
so the same \IdrisBound{gamma} is used everywhere.

\begin{centertight}
\begin{minipage}{0.35\textwidth}
\inferenceApp
\end{minipage}\hfill
\begin{minipage}{0.55\textwidth}
\ExecuteMetaData[Code/MiniVelo.idr.tex]{inferenceApp}
\end{minipage}
\end{centertight}

Finally, we have a rule where the premise's context has been extended:
a function of type (\TyFunc{a}{b}) is built by providing a term
of type $b$ defined in a context extended with a new variable of type $a$.

\begin{centertight}
\begin{minipage}{0.35\textwidth}
\inferenceFunc
\end{minipage}\hfill
\begin{minipage}{0.55\textwidth}
\ExecuteMetaData[Code/MiniVelo.idr.tex]{inferenceFunc}
\end{minipage}
\end{centertight}

Using this intrinsically typed representation, we can readily represent
entire typing derivations.
The following example\footnote{\IdrisData{Here} will be defined in
Section~\ref{sec:design:deBruijn} as a constructor for the \IdrisType{Elem} family.}
presents the internal representation
\IdrisFunction{Plus2} of the derivation proving that
$\exprLam{\exprInc{\exprInc{x}}}$ can be
assigned the type ($\TyFunc{\TyNat}{\TyNat}$).

\begin{centertight}
\begin{minipage}{0.4\textwidth}
\infer
  {\infer{\vdots}{\epsilon\,, \ty{x}{\TyNat} \vdash \ty{\exprInc{\exprInc{x}}}{\TyNat}}}
  {\epsilon \vdash \ty{\exprLam{\exprInc{\exprInc{x}}}}{\TyFunc{\TyNat}{\TyNat}}}
\end{minipage}\hfill
\begin{minipage}{0.5\textwidth}
\ExecuteMetaData[Code/MiniVelo.idr.tex]{Plus2Def}
\end{minipage}
\end{centertight}

By using \IdrisType{Term} as an \ac{ir} in our compiler
we have made entire classes of invalid programs unrepresentable:
it is impossible to form an ill scoped or ill typed term.
Indeed, trying to write an ill scoped or an ill typed program leads to a
static error as demonstrated by the following \IdrisKeyword{failing} blocks.%
\footnote{\Idris{} only accepts failing blocks if checking
their content yields an error matching the given string.}
In this first example we try to refer to a variable in an empty context.
\Idris{} correctly complains that this is not possible.

\ExecuteMetaData[Code/MiniVelo.idr.tex]{IllScoped}

In this second example we try to type the identity function as a function
from \TyNat to \TyBool. This is statically rejected as nonsensical:
\IdrisData{TyNat} and \IdrisData{TyBool} are distinct constructors!

\ExecuteMetaData[Code/MiniVelo.idr.tex]{IllTyped}

\noindent
Using such intrinsically typed \acp{edsl} we can statically enforce that
our elaborators do check that the raw terms obtained by parsing user input
are well scoped and well typed.
Writing our compiler passes (model-to-model transformations) and
evaluation engine (model-to-host transformation) using these
invariant-rich \acp{ir} additionally ensures that each step respects
the language's static semantics.
In fact we will describe in \Cref{sec:semantics} how we can use our \acp{edsl}
to both verify our static semantics whilst describing our dynamic semantics.

For languages equipped with more advanced type systems, that cannot be as easily
enforced statically, we can retain some of these guarantees by using a well
scoped core language rather than a well typed one.
This is the approach used in \Idris{} and it has already helped eliminate an
entire class of bugs arising when attempting to solve a metavariable with a
term that was defined in a different context~\cite{DBLP:conf/ecoop/Brady21}.

\subsection{Efficient \DeBruijn{} Representation}
\label{sec:design:deBruijn}

A common strategy for implementing well-scoped terms is to use typed
\emph{\DeBruijn{}} indices~\cite{MANUAL:journals/math/debruijn72},
which are easily realised as an inductive family~\cite{DBLP:journals/fac/Dybjer94}
indicating where in the type-level context the variable is bound.

Concretely, we index the \IdrisType{Elem} family by a context
(once again represented as a \IdrisType{SnocList} of kinds) and
the kind of the variable it represents.

\begin{centertight}
\begin{minipage}{0.10\textwidth}
\varRule
\end{minipage}\hfill
\begin{minipage}{0.80\textwidth}
\ExecuteMetaData[Code/MiniVelo.idr.tex]{ElemDecl}
\end{minipage}
\end{centertight}

We then match each context membership inference rule to a constructor.
The \IdrisData{Here} constructor indicates that the variable of interest is
the most local one in scope (note the non-linear occurrence of (\ty{x}{a}) on
the left hand side, and correspondingly of \IdrisBound{ty} on the right).

\begin{centertight}
\begin{minipage}{0.35\textwidth}
  \varZero
\end{minipage}\hfill
\begin{minipage}{0.55\textwidth}
\ExecuteMetaData[Code/MiniVelo.idr.tex]{varZero}
\end{minipage}
\end{centertight}

The \IdrisData{There} constructor skips past the most local variable to look for the variable of interest deeper in the context.

\begin{centertight}
\begin{minipage}{0.35\textwidth}
  \varSuc
\end{minipage}\hfill
\begin{minipage}{0.55\textwidth}
\ExecuteMetaData[Code/MiniVelo.idr.tex]{varSuc}
\end{minipage}
\end{centertight}

Whilst a valid definition, this approach unfortunately does not scale to
large contexts:
every \IdrisType{Elem} proof is linear in the size of the \DeBruijn{}
index that it represents.
To improve the runtime efficiency of the representation we instead opt to
model \DeBruijn{} indices as natural numbers, which \Idris{} compiles to
GMP-style unbounded integers.
Further, we need to additionally define an \IdrisType{AtIndex} family to ensure that
all of the natural numbers we use correspond to valid indices.
We pointedly reuse the \IdrisType{Elem} names because these \IdrisData{Here}
and \IdrisData{There} constructors play exactly the same role.

\ExecuteMetaData[Code/MiniDeBruijn.idr.tex]{AtIndexDef}

\noindent
We then define a variable as the pairing of a natural number and an \emph{erased}
(as indicated by the \IdrisKeyword{0} annotation on the binding site for \IdrisBound{prf})
proof that the given natural number is indeed a valid \DeBruijn{} index.

\ExecuteMetaData[Code/MiniDeBruijn.idr.tex]{IsVarDef}

Thanks to Quantitative Type Theory~\cite{DBLP:conf/birthday/McBride16,DBLP:conf/lics/Atkey18}
as implemented in \Idris{}, the compiler knows that it can safely erase these
runtime-irrelevant proofs.
we now have the best of both worlds: a well-scoped realisation of \DeBruijn{} indices
that is compiled efficiently.

\todo{Talk about smart constructors \& views?}

Just like the naïve definition of \DeBruijn{} indexing is not the
best suited for a practical implementation,
the inductive family \IdrisType{Term} described in Section~\ref{sec:design}
is not the most convenient to use.
We will now see one of its limitations and how we remedied it in
\Velo{}.


\subsection{Compact Constant Folding}
\label{sec:design:constants}

Software Foundations' \emph{Programming Language Foundations}
opens with a constant-folding transformation exercise~\cite[Chapter~1]{Pierce:SF2}.
Starting from a small language of expressions (containing natural numbers, variables, addition, subtraction, and multiplication) we are to deploy the semiring properties to simplify expressions.
The definition of the simplifying traversal contains much duplicated code due to the way the source language is structured:
all the binary operations are separate constructors, whose subterms need to be structurally simplified before we can decide whether a rule applies.
The correction proof has just as much duplication because it needs to follow the structure of the call graph of the function it wants to see reduced.
The only saving grace here is that Coq's tactics language lets users write scripts that apply to many similar goals thus avoiding duplication in the source file.

In \Velo{}, we structure our core language's representation in an algebraic
manner so that this duplication is never needed.
All builtin operators (from primitive operations on builtin types to function
application itself) are represented using a single \IdrisData{Call} constructor
which takes an operation and a type-indexed list of subterms.

\ExecuteMetaData[Code/MiniCompact.idr.tex]{TermDef}

Here \IdrisType{Terms} is the pointwise lifting of \IdrisType{Term} to lists
of types. In practice we use the generic \IdrisType{All} quantifier, but this
is morally equivalent to the specialised version presented below:

\ExecuteMetaData[Code/MiniCompact.idr.tex]{AllDef}

The primitive operations can now be enumerated in a single datatype
\IdrisData{Prim} which lists the primitive operation's arguments and
the associated return type.


\ExecuteMetaData[Code/MiniCompact.idr.tex]{PrimDef}

Using \IdrisType{Prim}, structural operations can now be implemented by handling recursive calls on the subterms of \IdrisData{Call} nodes uniformly before dispatching on the operator to see whether additional simplifications can be deployed.
Similarly, all of the duplication in the correction proofs is factored out in a single place where the induction hypotheses can be invoked.


\subsection{Well-Typed Holes}
\label{sec:design:holes}

Holes are a special kind of placeholder that programmers can use for parts of the program they have not yet written.
In a typed language, each hole will be assigned a type based on the context it is used in.

\emph{Type-Driven Development}~\cite{DBLP:journals/pacmpl/OmarVCH19}
is a practice by which the user enters into a dialogue
\emph{with} the compiler to interactively build the program.
We can enable type-driven programming in part by providing special language support for holes and operations on them.
Such operations will include the ability to inspect, refine, compute with, and instantiate (with an adequately typed term) holes.
We believe that bare-bones language support for type-driven development
should at least include the ability to:
(1) inspect the type of a hole and the local context it appears in;
(2) instantiate a hole with an adequately typed term;
and as well
(3) safely evaluate programs that still contain holes.
\Velo{} provides all three.

\Idris{} elaborates holes as it encounters them by turning them into
global declarations with no associated definition.
Because of this design choice users cannot mention the same hole explicitly in different places to state their intention that these yet unwritten terms ought to be the same.
Users can refer to the hole's solution by its name,
but that hole is placed in one specific position
and it is from that position that \Idris{} infers its context.

In \Velo{}, however, we allow holes to be mentioned arbitrarily many times in
arbitrarily different local contexts.
In the following example, the hole \texttt{?h} occurs in two distinct contexts:
$\epsilon,\,a,\,x$ and $\epsilon,\,a,\,y$.

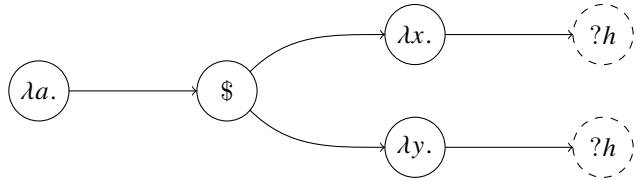
\begin{center}
  \holeexamplegraph{}
\end{center}

As a consequence, a term will only fit in that hole if it happens to live in the shared common prefix of these two contexts ($\epsilon,\,a$).
Indeed, references to $x$ will not make sense in $\epsilon,a,y$ and vice-versa for $y$.

Our elaborator proceeds in two steps.
First, a bottom-up pass records holes as they are found and, in nodes with multiple subterms, reconciles conflicting hole occurrences by computing the appropriate local context restrictions.
This process produces a list of holes, their types, and local contexts,
together with a \IdrisType{Holey} term that contains invariants ensuring
these collected holes do fit in the term.
Second, a top-down pass produces a core \IdrisType{Term} indexed by the list of \IdrisType{Meta} (a simple record type containing the hole's name, the context it lives in, and its type).
Hole occurrences end up being assigned a thinning that embeds the metavariable's actual context into the context it appears in.
We discuss thinnings and their use in \Velo{} in Section~\ref{sec:compiler-pass}.

Although these intermediate representations are \Velo{}-specific, the technique
and invariants are general and can be reused by anyone wanting to implement
well-scoped holes in their functional \ac{dsl}.


\section{Compiler Passes}
\label{sec:compiler-pass}

\todo{comment on cost of thinning as inductive values?}

Now that our core language is well-scoped by construction, our compiler passes must also be shown to be scope-preserving.
This is not a new requirement, merely it makes concrete a constraint that used to be enforced informally.
More importantly we show, with our compiler passes, that model-to-same-model transformation of our \ac{edsl} is possible, and that the infrastructure required is not bespoke to \Velo{}.

The purpose of \ac{cse} is to identify subterms that appear multiple times in the syntax tree, and to abstract over them to avoid needless recomputations at runtime.
In the following example for instance, we would like to let-bind $t$ before the application node (denoted \$) so that $t$ may be shared by both subtrees.

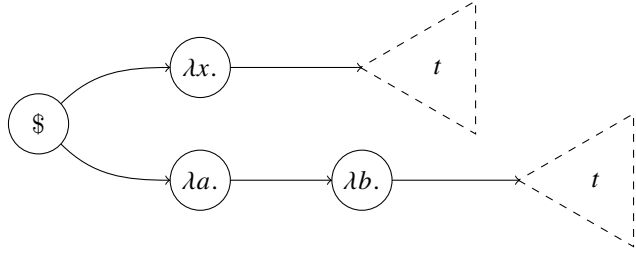
\begin{center}
  \cseexamplegraph{}
\end{center}

One of the challenges of \ac{cse}, as exemplified above, is that the term of interest may be buried deep inside separate contexts.
In our intrinsically scoped representation, $t$ in scope
$\Gamma,\, \ty{x}{\sigma}$ is potentially not actually
syntactically equal to a copy living in $\Gamma,\, \ty{a}{\tau},\, \ty{b}{\nu}$.
Indeed a variable $v$ bound in $\Gamma$ will, for instance, be
represented by the \DeBruijn{} index ($1+v$) in $\Gamma,\, \ty{x}{\sigma}$
but by the index ($2+v$) in $\Gamma,\, \ty{a}{\tau},\, \ty{b}{\nu}$.

If only terms were indexed by their exact \emph{support}
(i.e. a context restricted to the variables actually used in the term)!
We would not care about additional yet irrelevant variables that happen to be in scope.
The principled solution here is to switch to a different representation when performing \ac{cse}.
The co-\DeBruijn{} representation~\cite{DBLP:journals/corr/abs-1807-04085} provides exactly this guarantee.


In the co-\DeBruijn{} representation, every term is precisely indexed by its exact support.
That is to say that every subterm explicitly throws away the bound variables that are not mentioned in it.
By the time we reach a variable node, a single bound variable remains in scope:
precisely the one being referred to.

This process of throwing unused variables away is reified using thinnings
i.e. renamings that are injective, and order preserving.
We can think of thinnings as sequences of 0/1 bits, stating whether each variable
is kept or dropped.

Below, we give a graphical presentation (taken from~\cite{MANUAL:draft/Allais22})
of the $S$ combinator (the lambda term $\lambda g. \lambda f. \lambda x. g x (f x)$)
in co-\DeBruijn{} notation.
In it we represent thinnings (i.e. lists of bits) as lists of either
$\bullet$ (1) or $\circ$ (0).

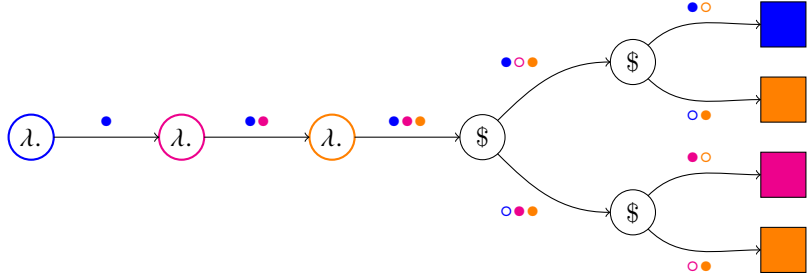
\begin{center}
  \codebruijnexamplegraph{}
\end{center}

The first three $\lambda$ abstractions only use $\bullet$ in their
thinnings because all of $g$, $f$, and $x$ do appear in the body of the combinator.
The first application node then splits the context into two:
the first subterm ($g x$) drops $f$
while the second ($f x$) gets rid of $g$.
Further application nodes select the one variable still in scope for each leaf subterm: $g$, $x$, $f$, and $x$.

Using a co-\DeBruijn{} representation, we can identify shared subterms:
they need to not be mentioning any of the most local variables and be syntactically equal.
Our pass successfully transforms the program on the left-hand side to the one
on the right-hand side where the repeated expressions \texttt{(add m n)}
and \texttt{(add n m)} have been let-bound.

\begin{center}
  \begin{minipage}[t]{0.4\textwidth}
    \begin{Verbatim}
let m = zero in
let n = (inc zero)
in (add (add (add m n) (add n m))
        (add (add n m) (add m n)))
    \end{Verbatim}
  \end{minipage}\hfill\begin{minipage}[t]{0.4\textwidth}
    \begin{Verbatim}
let m = zero       in
let n = (inc zero) in
let p = (add n m)  in
let q = (add m n)
in (add (add q p) (add p q))
    \end{Verbatim}
  \end{minipage}
\end{center}

This pass relies on the ability to have a compact representation of thinnings
(as the co-\DeBruijn{} representation makes heavy use of them),
and additionally the existence of a cheap equality test for them.
This is not the case in the implementation we include in \Velo{} but
it is a solved problem~\cite{MANUAL:draft/Allais22}.

\section{Execution}
\label{sec:semantics}

The \Velo{} \acs*{repl} lets users reduce terms down to head-normal forms.
We can realise \Velo{}'s dynamic semantics either through definitional
interpreters~\cite{10.1145/3093333.3009866,Augustsson1999edt},
or by providing a more traditional syntactic proof of
type-soundness~\cite{DBLP:journals/iandc/WrightF94}
but mechanised~\cite[Part 2: Properties]{plfa22.08} using inductive families.

We chose the latter approach: by using inductive families, we can make explicit
our language's operational semantics.
This enables us to study its meta-theoretical properties and in particular prove
a progress result: every term is either a value or can take a reduction step.
By repeatedly applying the progress result, until we either reach a value or the end
user runs out of patience and kills the process, this proof freely gives us an
evaluator that is guaranteed to be correct with respect to \Velo{}'s operational
semantics.

Following existing approaches~\cite[Part 2: Properties]{plfa22.08}, we have defined
inductive families describing how terms reduce.

\ExecuteMetaData[Code/MiniExecute.idr.tex]{ReduxDef}

As can be seen above, our setting enforces call-by-value:
as described by the rule \IdrisData{ReduceFuncApp}
(\exprApp{\exprLam{b}}{t}) only reduces to
($b \, \lbrace x \leftarrow t \rbrace$)
if $t$ is already known to be a value.
Furthermore, our algebraic design (\Cref{sec:design:constants}) allows
us to easily enforce a left-to-right evaluation order by having a generic
family describing how primitive operations' arguments reduce.
As can be seen below: when considering a type-aligned list of arguments,
either the \IdrisBound{hd} takes a step and the \IdrisBound{rest} is unchanged,
or the \IdrisBound{hd} is already known to be a value and a further argument
is therefore allowed to take a step.

\ExecuteMetaData[Code/MiniExecute.idr.tex]{ReducesDef}

We differ, however, from standard approaches by making our proofs of progress generic
such that the boilerplate for computing the reflexive transitive closure
when reducing terms is tidied away in a shareable module.
Our top-level progress definition is thus parameterised by reduction and value definitions:

\ExecuteMetaData[Code/MiniExecute.idr.tex]{ProgressDef}

\noindent
and the result of execution, which is similarly parameterised, is as follows
(where \IdrisType{RTList} is the type taking a relation and returning its
reflexive-transitive closure):

\ExecuteMetaData[Code/MiniExecute.idr.tex]{ResultDef}

The benefit of our approach is that language designers need only provide details of
what reductions are,
and how to compute a single reduction, the rest comes for free.
Moreover, with the result of evaluation we also get the list of reduction steps made that can, optionally, be printed to show a trace of execution.


%
%

\section{Conclusion}
\label{sec:conclusion}

We have shown that dependently typed languages satisfy the core requirements from the \emph{Language Workbench Challenge}~\cite{DBLP:conf/sle/ErdwegSVBBCGHKLKMPPSSSVVVWW13}.
\Velo{}'s \emph{notation} as a \ac{dsl} is, by design, textual, and the internal core bounded by \Idris{}'s own notation requirements.
More importantly the \emph{semantics} (statics and dynamics) of \Velo{}
are verified as part of the implementation thanks to the dependently typed setting.
The weakest supported core criteria, unfortunately, is that for \emph{editor} support.
Languages created through \Idris{} do not get an editor, they are free form languages which require their parsers and elaborators be hand written.
This can change with future investigation.
\Idris{} has support for elaborator reflection~\cite{DBLP:conf/icfp/ChristiansenB16} which provides a vehicle through which deriving parsers and elaborators can happen.

There are, however, more criteria from the language workbench feature model to consider:
semantic \& syntactic services for editors;
testing \& debugging;
and
composability.

With the rise of the \ac{lsp} it would be a good idea to look at how we can derive \ac{lsp} compatible language servers generically, thus addressing the missing provision of the optional semantic and syntactic services.
\Idris{} itself provides an \emph{IDE-Protocol}
, and there is support for the \ac{lsp} in \Idris{}.

Our languages also do not come with the ability to test and debug their
implementation.
Some of the features we have presented are fully formalised (e.g.\ execution),
others are only known to be scope-and-type preserving (e.g.\ \ac{cse}).
Therefore the dependently typed setting does not mean we do not need testing
anymore.
Prior work on generators for inductive families~\cite{DBLP:journals/pacmpl/LampropoulosPP18}
should allow us to bring property-based testing~\cite{DBLP:conf/icfp/ClaessenH00} to our core passes.

Finally there is language composability.
%
%
It would be advantageous to support the reuse of existing languages,
and their type-systems when designing new ones.
This is a hard problem:
One has to not only combine their semantics but also the remainder of the workbench tooling.
The \emph{language fragments} approach~\cite{10.1145/3563355} provides a
solution to language composability for intrinsically typed definitional
interpreters, but this does not extend to workbench tooling.
Extending this approach to our definition of semantics based on inductive
families and to creating composable workbench tooling is an open problem.

We strongly believe that with future engineering we can satisfy these missing criteria,
and make dependently typed languages a mighty fine language workbench.



\newpage
\bibliography{paper}

%
%
%

\end{document}